\documentclass[twocolumn,showpacs,preprintnumbers,amsmath,amssymb]{revtex4}

\usepackage{graphicx}
\usepackage{dcolumn}
\usepackage{bm}

\begin{document}

\preprint{APS/123-QED}

\title{Deformed Special Relativity with an invariant minimum speed as an explanation of the cosmological constant}  
\author{Cl\'audio Nassif Cruz}

\altaffiliation{{\bf UFOP}: Universidade Federal de Ouro Preto, Morro do Cruzeiro, Bauxita, 35.400-000-Ouro Preto-MG, Brazil.\\
e-mail: claudionassif@yahoo.com.br}

\date{\today}

\begin{abstract}
 This paper shows the need of the emergence of a universal minimum speed in the space-time by means of a more thorough investigation of Dirac's large number hypothesis (LNH). We will realize that there should be a minimum speed $V$ with the same status of the invariance of the speed of light $c$, however $V$ has gravitational origin. Hence, such a minimum speed forms a new kinematic basis in the space-time, thus leading to a new Deformed Special Relativity (DSR) for the quantum world so-called Symmetrical Special Relativity (SSR). Furthermore, we will show that such a new structure of space-time (SSR) reveals a connection between $V$ and a preferred reference frame $S_V$ of background field that leads to the cosmological constant $\Lambda$, which can be associated with a cosmological anti-gravity. We will also investigate the effect of the number of baryons $N$ (Eddington number) of the observable universe on the Hydrogen atom. Finally, we will show that SSR-metric plays the role of a de-Sitter (dS)-metric with a positive cosmological constant, which could assume a tiny value. 
\end{abstract}

\maketitle

\section{\label{sec:level1} Introduction}

We will search for a new structure of space-time with the presence of a minimum speed $V$ that behaves like a kinematic invariant for particles with low energies as is the speed of light $c$ for high energies. In order to do that, we will start from a deeper investigation of Dirac's large number hypothesis (LNH), where the relationship of the electric and gravitational interactions becomes more evident in the sense that such investigation of LNH will allow us to perceive that there should be a kinematic aspect of gravity represented by the invariant minimum speed $V(\propto G^{1/2}/\hbar)$. So we are led to think that such symmetry due to $c$ and $V$ forms the kinematic basis in a space-time that bahaves like a de-Sitter (dS) space-time represented by a positive cosmological constant $\Lambda$. In view of such a connection between Deformed Special Relativity (DSR) with a minimum speed and a DS space-time, we could get a tiny value of $\Lambda$, which is associated to a cosmological anti-gravity. 

The search for understanding the origin of the vacuum energy density $\rho$ related to $\Lambda>0$ in the scenario of an accelerated expanding universe has been the issue of hard investigations\cite{1}\cite{2}, where it is known that $\rho=\Lambda c^2/8\pi G$. 

The fine structure constant $\alpha$ is associated with the cosmological constant $\Lambda$\cite{hgn,hgn2,hgn3}. Thus, a possible variation of the fine structure constant\cite{sergio,sergio1,sergio2,pad,pad2} would also point to a fundamental change in the subatomic structure, since $\alpha$ has a property of connecting the micro and macro-world, whose age is measured by the speed of light $c$, i.e., $R_H=cT_H$, where $T_H(\cong 13.7$ Gyear) is the Hubble time and $R_H(\sim 10^{26}m)$ is the Hubble radius, i.e., the radius of the visible universe.  

The relationship between $\alpha$ and $\Lambda$, which is linked to the dark sector of the universe is associated with models that aim to explain the anti-gravitational effects of the dark energy based 
on scalar fields
\cite{cine1,cine2,cine3,cine4,cine5,cine6,
cine7}. 

The emergence of a minimum speed $V$ in the
space-time is associated with a preferred reference frame, thus leading to the birth of a new relativity with Lorentz symmetry violation at lower energies, i.e., the so-called Symmetrical Special Relativity (SSR)\cite{N2016,N2015,N2012,N2010,N2007,Rodrigo,Rodrigo2,Rodrigo3,N2018,uncertainty}.

It has also been shown that SSR has a relationship with the principle of Mach\cite{Rodrigo3,mach,mach2,mach3} within a quantum scenario due to the presence of the vacuum energy\cite{Rodrigo}.

The relationship between the fine structure constant and the cosmological constant is $\Lambda\propto\alpha^{-6}$\cite{hgn,hgn2,hgn3}. Actually, 
$\Lambda$ is also associated with other constants such as the mass of the electron $m_{e}$, Planck constant ($\hbar$) and the universal constant of gravity $G$. Thus, we can realize that $\Lambda$ is connected to the constants of the standard model of elementary particles, namely $\Lambda\sim (G^2/\hbar^4)(m_e/\alpha)^6$\cite{hgn,hgn2,hgn3}.

We will investigate the global effect of the number of baryons $N$ of the universe (Eddington number) on the Hydrogen atom within a Machian scenario
\cite{mach,mach2,mach3}. 

The last section will be dedicated to the introduction of the space-time and velocity transformations for ($1+1$)D in SSR-
theory\cite{N2016,N2015,N2012,N2010,N2007,Rodrigo,Rodrigo2,Rodrigo3}. 

Our main goal is to show that the SSR-metric plays the role of a dS-metric, so that $\Lambda$ emerges naturally from the SSR-theory. Thus, $\Lambda$ is related to a cosmological anti-gravity. The tiny value of $\Lambda$ could be estimated. 
 
\section{\label{sec:level1} A minimum speed as a universal constant}
 
First of all, we intend to show the existence of a universal minimum speed $V$ as a new fundamental constant of nature. Such a speed has the same status of the invariance of the speed of light $c$\cite{N2016}, however $V$ is given for lower energies. We will show the emergence of $V$ within the Dirac's LNH scenario. 

The relationship between the minimum speed $V$ and the cosmological constant $\Lambda$ should be understood by means of the so-called ultra-referential $S_V$, i.e., a preferred reference frame shown 
in Fig.1\cite{N2016,N2015,N2012,N2010,N2007,Rodrigo,Rodrigo2}, which is associated with $V$. So $V$ is related to a background field of gravitational origin as foundation of the cosmological constant $\Lambda$\cite{Rodrigo3,mach,mach2,mach3}. 

We will conclude that there should be an equivalence between SSR-metric and a de-Sitter (dS) metric with the presence of $\Lambda$ to be investigated later. 

In order to obtain the universal minimum speed $V$, we will start from Dirac's LNH that introduces the ratio of the electric and gravitational forces between the electron and proton in the Hydrogen atom\cite{hgn,hgn2,hgn3}, namely:

 \begin{equation}
 \frac{F_e}{F_g}=\frac{e^2}{Gm_pm_e}=\frac{q_e^2}{4\pi\epsilon_0Gm_pm_e},  
 \end{equation}
 where $F_e/F_g\sim 10^{40}$. The masses $m_e$ and $m_p$ are the electron and proton masses respectively. We write 
 $q_e^2/4\pi\epsilon_0=e^2$. 
 
 The large number of the order of $10^{40}$ is the well-known Dirac's large number. Such number coincides exactly with
 $\sqrt{N}(\sim 10^{40})$, where $N(\sim 10^{80})$ is the well-known Eddginton number, i.e., the number of protons (baryons) in the universe. 
 
 Is is interesting to note that such large number ($\sqrt{N}$) can be obtained by other ways, as for instance, the ratio $F_e/F_g\sim R_H/r_p\sim 10^{40}$, where $r_p$ is the proton radius and $R_H$ is the Hubble radius. This indicates that this large number connects the length scales of the quantum world (the proton radius) with the cosmological scale, i.e., the Hubble radius. 

 By using the theorem of work (energy) to provide the works of the electric and gravitational forces to ionize the Hydrogen atom and a hypothetical Hydrogen atom with only gravitational interaction between the proton and electron, which in turn would be taken from its Bohr radius $a_0$ to infinite, then we obtain the ratios of the works of both applied forces and their kinetic energies within the Dirac's LNH scenario in order to get the minimum speed $V$ of gravitational origin ($V\propto G^{1/2}/\hbar$), namely: 

\begin{equation}
 \frac{\mathcal W_{F_e(\infty\rightarrow a_0)}}{\mathcal W_{F_g(\infty\rightarrow a_0)}}=\frac{\frac{q_e^2}{4\pi\epsilon_0}\int_{\infty}^{a_0}\frac{1}{r^2}dr}
 {Gm_pm_e\int_{\infty}^{a_0}\frac{1}{r^2}dr}=\frac{F_e}{F_g}=\frac{\frac{1}{2}m_ev_B^2}{\frac{1}{2}m_eV^2}, 
\end{equation}
where $v_B(=e^2/\hbar\sim10^{5}m/s)$ is the speed of the electron in the fundamental state of the Hydrogen atom, which is a universal constant so-called Bohr velocity. It can be alternatively written as $v_B=\alpha c$, where the constant $\alpha=e^2/\hbar c(\approx 1/137$) is the well-known {\it fine structure constant}. 

On the other hand, the speed $V$ shown in Eq.(2) shoud be understood as the most fundamental speed related to a small classical kinetic energy, since it has origin from the work of the weakest force of nature (the gravitational force) as being the negative of the same applied force to ionize a hypothetical gravitational Hydrogen atom, where we just consider the gravitational interaction between the proton and electron that form the most stable and simple bind structure in the universe, i.e., the Hydrogen atom. 

Therefore, it does not seem by chance that Dirac's LNH with its strong ``coincidences'' of ratios between scales of space-time with the same order of magnitude of $10^{40}$ starts just from the Hydrogen atom as the fundamental (stable) structure that justifies the emergence of the universal minimum speed $V$ with the same status of the speed of light $c$. 

Furthermore, the extended Dirac's LNH given in Eq.(2) shows clearly a new fundamental symmetry with respect to the existence of a higher energy scale related to the Bohr velocity ($v_B=\alpha c$) with electric origin, i.e., a kinetic energy of Colombian origin to ionize the Hydrogen atom, namely $(1/2)m_ev_B^2$, and a lower (zero-point) energy related to a minimum speed $V$, which is associated with a minimum classical kinetic energy of ionization of a hypothetical Hydrogen atom by just considering the gravitational force between the mass of the proton ($m_p$) and electron ($m_e$). This leads to a very low kinetic energy $(1/2)m_eV^2$ of gravitational origin. 

With respect to the symmetry provided by the extended Dirac's LNH, we have a good reason for considering the speed $V$ as the kinematic invariant given for any particles that never reach $V$ at lower energies, $V$ being related to a preferred reference frame (Fig.1) associated with the vacuum energy that leads to the cosmological constant, as we will see later. This motivates us to build the so-called Symmetrical Special Relativity (SSR)\cite{N2007} for describing the motion of the particles in the quantum world with the presence of the vacuum energy associated with the ultra-referential $S_V$ (Fig.1). 

However, if we admit the classical idea of rest in the quantum world by making $V\rightarrow 0$, this would violate the Dirac's LNH, as gravity and the vacuum energy would vanish and thus the large number would diverge ($F_e/F_g\rightarrow\infty$), so that the universe also would be infinite ($N\rightarrow\infty$) with an infinite Hubble radius ($R_H\rightarrow\infty$). Therefore, we realize that SR is not consistent with the Dirac's LNH scenario, as rest ($v=0$) is naturally conceived by the classical theory, where there is no minimum speed, i.e., we just make $V=0$ in order to recover SR as a particular case of SSR. Thus, only SSR is consistent with Dirac's LNH scenario.   

Actually, we should realize that the extended Dirac's LNH shows that the universe would be inconceivable if there is no minimum speed or if rest were possible to conceive in the quantum world, since the existence of a non-null minimum speed $V$ (a zero-point energy) in the quantum world is essentially responsible by gravity, thus leading to the kinematic basis of a quantum gravity at lower energies for describing the vacuum and its cosmological implications (the cosmological constant). 

By substituting the Bohr velocity $v_B(=e^2/\hbar)$ in Eq.(2) and, after by performing the calculations, we finally obtain $V$, namely: 

\begin{equation} 
V=\sqrt{\frac{Gm_em_p}{4\pi\epsilon_0}}\frac{q_e}{\hbar}, 
\end{equation}
where we write $e=q_e/\sqrt{4\pi\epsilon_0}$. 

From Eq.(3), we get $V\cong 4.58\times 10^{-14}m/s$. 

It is important to realize that the minimum speed $V$ is directly related to the Planck minimum length of quantum gravity ($L_P$), namely $V=\sqrt{Gm_pm_e}e/\hbar=(e\sqrt{m_pm_ec^3/\hbar^3})L_{P}=L_P/\tau$, where we obtain $L_P=\sqrt{G\hbar}/c^3\sim 10^{-35} $m and $\tau=(e\sqrt{m_pm_ec^3/\hbar^3})^{-1}\sim 10^{-21}$s, such that, if we consider $L_P\rightarrow 0$, this leads to 
$V\rightarrow 0$ and thus we recover the classical space-time of SR without quantum gravity effect. This leads us to perceive that there is a connection of the minimum speed $V$ with the Planck length ($L_P$) by means of the constant of gravity ($G$) and the Planck constant ($\hbar$), as we find $V\propto L_P\propto G^{1/2}$. So, both $V$ and $L_P$ are respectively the kinematic and space ingredients of quantum gravity, since $L_P$ is well-known as the fundamental length of quantum gravity. 

In Eq.(3), if we make $G\rightarrow 0$, we find $V\rightarrow 0$. But, as gravity does not vanish in anywhere, rest is in fact forbbiden due to a zero-point energy related to the fundamental vacuum energy. Hence, we are led to postulate $V$ as a kinematic invariant connected to the fundamental vacuum at very low energies. 

By combining Eq.(2) given for the Dirac's LNH with Eq.(3), we alternatively obtain 

\begin{equation}
\frac{F_e}{F_g}=\frac{v_B^2}{V^2}=\frac{4\pi\epsilon_0v_B^2\hbar^2}{Gm_em_pq_e^2}=
\frac{q_e^2}{4\pi\epsilon_0Gm_em_p}, 
\end{equation}
which is in fact consistent with Eq.(2), as it is already known that 
$v_B=q_e^2/4\pi\epsilon_0\hbar=e^2/\hbar$. 

\subsection{\label{sec:level1} The effect of the Eddington Number on the Hydrogen atom}

From Eq.(4), we can write the most fundamental speed $V(\cong 1.5322\times 10^{-22}c)$ of gravitational origin in function of the Bohr velocity $v_B(\cong 137^{-1}c)$ of Coulombian origin, as follows: 

\begin{equation}
 V=\frac{\sqrt{4\pi\epsilon_0Gm_em_p}}{q_e}v_B, 
\end{equation}
where we have $v_B=q_e^2/4\pi\epsilon_0\hbar$ and $\xi=V/c=\sqrt{4\pi\epsilon_0Gm_em_p}v_B/q_e=\sqrt{Gm_em_p/4\pi\epsilon_0}q_e/\hbar c\cong 1.5322\times 10^{-22}$. The dimensionless constant $\xi$ represents the so-called {\it fine tuning constant}\cite{N2016} as being a very fine structure constant of gravito-electromagnetic origin. 

Now it is important to further clarify why the minimum speed $V$ must depend specially on the proton mass instead of other lighter particles, so that there should not be a minimum speed with a value smaller than $V$. Thus, for instance, if we think about a positronium atom, the mass of the proton in Eq.(5) would be replaced by the mass of the electron or positron, so that we get a minimum speed $V^{\prime}$ smaller than $V$. However, it must be emphasized that the positronium is highly unstable by decaying rapidly. If so, this would lead to a rapid collapse of the space-time. Due to such a collapse that would prevent the existence of the universe for a much 
longer time, it is easy to conclude that $V^{\prime}(<V)$ cannot be considered for our obvervable universe with the age of about $13.7$ Gyear. 
 
In summary, the only bound state of two particles that is lighter and also completely stable is the Hydrogen atom. This is the reason why we postulate 
$V$ as the universal minimum speed. 
 
Other examples of bound states can also be provided for verifying that only the Hydrogen atom has complete stability in its ground state, and it is simultaneously the lightest stable element with highly stable particles such as the proton and electron. 

From Eq.(3), Eq.(4) and Eq.(5), we can write 

\begin{equation}
v_B=\frac{q_e^2}{4\pi\epsilon_0\hbar}=\sqrt{\frac{F_e}{F_g}}V=
\frac{q_e}{\sqrt{4\pi\epsilon_0Gm_em_p}}V, 
\end{equation}
where $F_e/F_g\sim\sqrt{N}\sim 10^{40}$ is the Dirac`s large number, so that we have $\sqrt{F_e/F_g}\sim\sqrt[4]{N}\sim 10^{20}$ and
$N\sim 10^{80}$ as being the well-known Eddington number that represents the order of magnitude of number of baryons in the universe. Thus, we can write Eq.(6) as $v_B\sim\sqrt[4]{N}V$, which connects the quantum world (Hydrogen atom) to the cosmological quantity represented by the number $N$ of baryons in the universe and also the vacuum energy associated with the universal mininum speed $V$ related to the cosmological constant $\Lambda$. 
 
The numerical coefficient $\sqrt[4]{N}\sim 10^{20}$ is known as the Weyl number. This is the great evidence of the connection of local quantities (quantum quantities) with the cosmological quantities. Thus, we can also say that this is an achievement of a kind of Machian principle\cite{mach,mach2,mach3}, where the global distribution of mass (number $N$ of baryons) determines the local properties of atoms, as is in the case of the Hydrogen atom. 
 
We can also obtain the Bohr radius ($a_0$) in terms of cosmological quantities such as the Weyl number. Thus, knowing that $v_B=q^2_e/4\pi\epsilon_0\hbar\sim\sqrt[4]{N}V$ [Eq.(7)] and also $a_0=\hbar/m_ev_B$, we find the Bohr radius ($a_0$) in terms of $V$ and $N$, namely:
 
\begin{equation}
a_0(N)\sim\frac{\hbar}{\sqrt[4]{N}m_eV}=\frac{\hbar^2}{\sqrt[4]{N}\sqrt{Gm_e^3m_p}e}, 
\end{equation}
where we have considered $V=\sqrt{Gm_em_p}e/\hbar$, with $e=q_e/\sqrt{4\pi\epsilon_0}$ [Eq.(3)]. 

If we make $N=1$ in Eq.(7), we would find a giant Hydrogen atom with a radius in the order of magnitude of about ten ($10$) stars like the sun, i.e, $a_0(N=1)\sim 10^{10}$m.

If $N\rightarrow 0$, we would have no Hydrogen atom, i.e., there would not be baryons in the universe, so that we would have just an enourmous quark condensate for representing the universe. This would be similar to a giant collapsed star made only of quarks. 

If $N\rightarrow\infty$, we also would have no Hydrogen atom, because it would be completely crushed, so that $a_0\rightarrow 0$. This case shows us clearly how the entire mass of the universe represented by the number $N$ is able to locally compress the Hydrogen atom due to a kind of ``pressure" of Machian (non-local) origin that would go to infinite if $N$ diverges by reducing the Bohr radius to zero. Here we might also think about the emergence of a singularity due to the formation of a completely collapsed structure. 

In view of the dependence of the Bohr velocity $v_B$ with the Weyl number ($\sqrt[4]{N}\sim 10^{20}$) or even the Eddginton number $N$, we also should have a dependence of the fine structure constant 
$\alpha$ with $N$, which can be obtained from Eq.(7), as follows:

\begin{equation}
\alpha(N)=\frac{v_B}{c}\sim\sqrt[4]{N}\frac{V}{c}=\sqrt[4]{N}\xi=
\sqrt[4]{N}\sqrt{Gm_em_p}\frac{e}{\hbar c},
\end{equation}
where we have $\xi=V/c=\sqrt{Gm_em_p}e/\hbar c$\cite{N2016} and 
$V=\sqrt{Gm_em_p}e/\hbar$, with $e=q_e/\sqrt{4\pi\epsilon_0}$. 

\section{\label{sec:level1} Space-time and velocity transformations in SSR} 

In this section, first of all we will carefully investigate the concepts of reference frame in SSR and their implications such as the new transformations of space-time and velocity.

We will also show the equivalence between the SSR-metric and a dS-metric, which is associated with a certain positive cosmological constant 
$\Lambda(>0)$, thus leading to a cosmological anti-gravity. So, we will conclude that SSR generates
a background metric that plays the role of a dS-metric with the presence of $\Lambda$. In order 
to realize such fundamental equivalence, we will
use a toy model as it will be presented in this 
section. 

Violation of Lorentz symmetry for very low energies\cite{N2016} generated by the presence of a background field related to $S_V$ (Fig.1) creates a new causal structure of space-time with an invariant mimimum speed $V=\sqrt{Gm_pm_e}e/\hbar$ [Eq.(3)], which is an unattainable limit of speed for all particles at lower energies. 

Since the minimum speed $V$ is an invariant quantity as is the speed of light $c$, $V$ does not alter the speed $v$ of any particle, as we will show later. Therefore, we denominate ultra-referential $S_V$ as being the preferred reference frame in relation to which we have the speeds $v$ of any particles (Fig.1). In view of this, the well-known Lorentz transformations are changed in the presence of the background reference frame $S_V$ (Fig.1). 

In the case $(1+1)D$ (Fig.1), we obtain the space-time transformations given between the running reference frame $S^{\prime}$ and the background reference frame $S_V$, namely: 

\begin{figure}
\begin{center}
\includegraphics[scale=0.09]{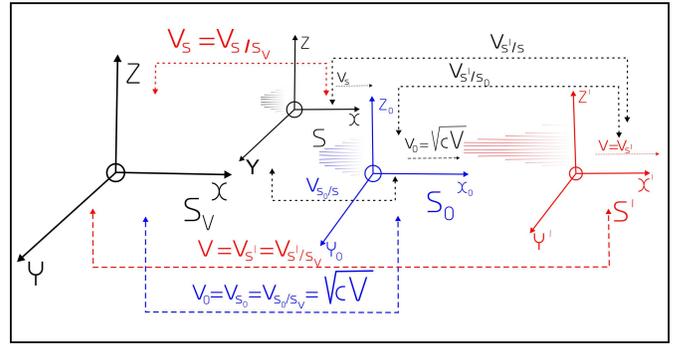} 
\end{center}
\caption{The reference frame $S^{\prime}$ moves in $x$-direction with a speed $v(>V)$ with respect to the universal reference frame, i.e., the ultra-referential $S_V$ associated with $V$. In this figure, we see the two running referentials $S$ and $S^{\prime}$ with speeds $v=v_S=v_{S/S_V}$ and $v^{\prime}=v_S'=v_{S'/S_V}$, both of them given in relation to the background frame (ultra-referential $S_V$), plus two fixed referentials $S_0$ with speed $v_0=v_{S_0/S_V}=\sqrt{cV}$ given with respect to the background frame $S_V$ and the own ultra-referential $S_V$ of vacuum associated with the unattainable minimum speed $V$. Thus we can find the relative velocity between $S'$ and $S$, i.e., $v_{rel}=v_{S'/S}$, which is shown clearly in Eq.(13) and Eq.(14), thus leading to some important cases, as for instance: a) If only the running referential $S$ coincides with $S_0$ ($S\equiv S_0$), we find the relative velocity between $S'$ and $S_0$, i.e., $v_{rel}=v_{S'/S_0}$. b) If only the running referential $S'$ coincides with $S_0$ ($S'\equiv S_0$), we find the relative velocity between $S_0$ and $S$, i.e., $v_{rel}=v_{S_0/S}$, as it is also indicated in this figure.} 
\end{figure}

\begin{equation}
dx^{\prime}=\frac{\sqrt{1-V^2/v^2}}{\sqrt{1-v^2/c^2}}[dX-v(1-\alpha)dt]
\end{equation}

and 

\begin{equation}
 dt^{\prime}=\frac{\sqrt{1-V^2/v^2}}{\sqrt{1-v^2/c^2}}\left[dt-\frac{v(1-\alpha)dX}{c^2}\right], 
 \end{equation}
with $\alpha=V/v$ and $\Psi=\theta\gamma=\sqrt{1-V^2/v^2}/\sqrt{1-v^2/c^2}$, where $\theta=\sqrt{1-V^2/v^2}$ and $\gamma=1/\sqrt{1-v^2/c^2}$. 

The coordinates $X$ shown in the transformations $(1+1)D$ above, $Y$ and $Z$ (Fig.1) form the ultra-referential $S_V$ connected to the vacuum energy. 

The inverse transformations for this special case $(1+1)D$ (Fig.1) were demonstrated in a previous work\cite{N2016}. Of course, if we make $V\rightarrow 0$, we recover the Lorentz transformations. 

The general transformations in the space-time $(3+1)D$ of SSR were also shown in a previous paper\cite{N2016}. In an another previous paper\cite{N2015}, it was first shown that SSR transformations breaks down the Lorentz and Poincar\'e's groups.

This new causal structure of space-time, i.e., the Symmetrical Special Relativity (SSR) presents the following energy $E$ and momentum $P$ for a particle,  namely $E=m_0c^2\Psi=m_0c^2\sqrt{1-V^2/v^2}/\sqrt{1-v^2/c^2}$\cite{N2016}\cite{N2007}, in such a way that $E\rightarrow 0$ when $v\rightarrow V$, and $P=m_0v\Psi=m_0v\sqrt{1-V^2/v^2}/\sqrt{1-v^2/c^2}$\cite{N2016}\cite{N2007}, such that $P\rightarrow 0$ when $v\rightarrow V$. 

It is important to notice that both momentum-energy of a particle in SSR is $P_0=m_0v_0=m_0\sqrt{cV}$ and $E_0=E(v_0)=m_0c^2$ for $v=v_0=\sqrt{cV}\neq 0(>V)$, as we find $\Psi(v_0)=\Psi(\sqrt{cV})=1$, where the energy $m_0c^2$ is exactly equivalent to the rest energy in SR, since there is no rest in SSR. This means that the momentum never vanishes in SSR due to the invariant minimum speed $V$, as there is an intermediary speed $v_0(=\sqrt{cV})$ given with respect to the preferred frame $S_V$ (Fig.1), so that the momentum is non-null ($P_0$) and the energy is equivalent to the rest energy $m_0c^2$ in SR with $p=0$ for $v=0$. However, in SSR, $E_0$ is associated with the speed $v_0$ (reference frame $S_0$) with respect to the ultra-referential $S_V$, as there is no rest in the space-time of SSR, where the references frames $S$, $S^{\prime}$ and specially $S_0$ for $v=v_0(=\sqrt{cV})$ are shown in Fig.1. 

The SSR-metric is a deformed Minkowski metric with the presence of the multiplicative factor $\Theta=\Theta(v)=1/(1-V^2/v^2)$\cite{Rodrigo}, which plays the role of a conformal factor as already shown in a previous paper\cite{Rodrigo}, thus leading to a Conformal Special Relativity represented by SSR due to the presence of the invariant minimum speed $V$, as follows: 

\begin{equation}
d\mathcal S^{2}=\frac{1}{\left(1-V^2/v^2\right)}[c^2(dt)^2-(dx)^2-(dy)^2-(dz)^2],
\end{equation}
or simply $d\mathcal S^{2}=\Theta\eta_{\mu\nu}dx^{\mu}dx^{\nu}$, where $\Theta=1/(1-V^2/v^2)$ and $\eta_{\mu\nu}$ is the Minkowski metric. 

By dividing Eq.(9) by Eq.(10), we obtain the following velocity transformation in SSR, namely: 

\begin{equation}
v_{rel}=v_{S'/S}=\frac{v^{\prime}-v+V}{1-\frac{v^{\prime}v}{c^2}+\frac{v^{\prime}V}{c^2}}=\frac{v^{\prime}-v(1-\alpha)}{1-\frac{v^{\prime}v(1-\alpha)}{c^2}}, 
\end{equation}
where $\alpha=V/v$.

We have considered $v_{rel}=v_{relative}=v_{S'/S}\equiv dx^{\prime}/dt^{\prime}$ and $v^{\prime}\equiv dX/dt$ when dividing Eq.(9) by Eq.(10). 

We should stress that $v^{\prime}=v_{S'}\equiv dX/dt$ is the motion of the referential $S^{\prime}$ (Fig.1) with respect to the background reference frame $S_V$ connected to the unattainable minimum speed $V$, i.e., we can write the notation $v_{S'}=v_{S'/S_V}$ for representing the absolute motion of $S^{\prime}$, which is observer-independent, as $S_V$ is absolute for being the preferred reference frame.  

The speed $v$ shown in Fig.1 represents the motion of the referential $S$ with respect to the background reference frame $S_V$, i.e., we can write the notation $v=v_{S}=v_{S/S_V}$ for representing the absolute motion of $S$ (Fig.1), which is also observer-independent.

The speed $v_{rel}$ is the relative speed between the absolute speeds $v_{S'}$ and $v_{S}$, both of them given in relation to the background framework $S_V$, i.e., we have $v_{rel}=v_{S'/S}$ (Fig.1). So we can rewrite the speed transformation in Eq.(12) by using the notations with the presence of the background frame $S_V$, namely: 

\begin{equation}
v_{rel}=v_{S'/S}=\frac{v_{S'/S_V}-v_{S/S_V}+V}{1-\frac{(v_{S'/S_V})(v_{S/S_V})}{c^2}+\frac{(v_{S'/S_V})V}{c^2}}, 
\end{equation}
where $v=v_{S}=v_{S/S_V}$ (speed $v$ of the reference frame $S$ in relation to $S_V$) and $v'=v_{S'}=v_{S'/S_V}$ (speed $v^{\prime}$ of the reference frame 
$S^{\prime}$ in relation to $S_V$).  

Fig.1 also shows the reference frame $S_0$, whose speed $v_0(=\sqrt{cV})$ is also given in relation $S_V$.

As $v_0$ is the intermediary speed, such that $V<<v_0<<c$ with $\Psi(v_0)=\Psi(\sqrt{cV})=1$, all the speeds $v$ not so far from $v_0$, where $E\approx E_0=m_0c^2$ represent the Newtonian approximation within the scenario of SSR, as we get $\Psi(V<<v<<c)\approx 1$. 

If $V\rightarrow 0$, Eq.(13) would recover the Lorentz velocity transformation, where both speeds $v_{S'}$ and $v_S$ would be given simply in relation to a certain Galilean frame at rest in lab, such that the background frame $S_V$ would vanish and thus $v_0$ would be also zero, i.e., the reference frame $S_0$ would become simply a certain Galilean reference frame at rest in lab. 

From the transformation in Eq.(13), let us just consider the important cases, where we must consider $v_{S'}\geq v_{S}$ (Fig.1), namely: 

 {\bf a)} If $v_{S'}=c$ (photon) and $v_S\leq c$, this implies in $v_{rel}=c$. Such result just verifies the invariance of $c$.

 {\bf b)} If $v_{S'}>v_S(=V)$, this implies in $v_{rel}=``v_{S'}-V"=v_{S'}$. For example, if $v_{S'}=2V$ and $v_S=V$, this leads to $v_{rel}=``2V-V"=2V$, which means that $V$ ($S_V$) really has no influence on the speeds of any particles. Thus, $V$ works as if it were an ``absolute zero of motion'', being invariant and having the same value at all directions of space $3D$ of the isotropic background field associated with $S_V$. 

 {\bf c)} If $v_{S'}=v_S$, this implies in $v_{rel}=v_{S'/S}=``v_S-v_S"=``v-v"$($\neq 0)=\frac{V}{1-\frac{v_S^2}{c^2}(1-\frac{V}{v_S})}=
 \frac{V}{1-\frac{v^2}{c^2}(1-\frac{V}{v})}$. 

From the case ({\bf c}), let us consider two specific cases, as follows:

-$c_1$) Assuming that $v_S=V$, this implies in $v_{rel}=``V-V"=V$ as verified before. Indeed $V$ is an invariant minimum speed. 

-$c_2$) If $v_S=c$ (photon), this implies in $v_{rel}=c$, where we have the interval $V\leq v_{rel}\leq c$ given for the interval $V\leq v_S\leq c$. However, it must be stressed that there is no ordinary massive particle exactly at the ultra-referential $S_V$ with $v=v_S=V$. So, this is just a hypothetical condition to verify the consistency of the transformation in Eq.(13) with respect to the invariance of the minimum speed $V$, as already verified in the specific case $c_1$. 

This last case ({\bf c}) shows that it is impossible to find the rest for the particle on its own reference frame $S$, where $v_{rel}(v_S)$ ($\equiv\Delta v(v_S)$) is a function that increases with the increasing of $v=v_S$ of the referential $S$ (Fig.1). However, if we make $V\rightarrow 0$, so we would have $v_{rel}\equiv\Delta v=0$ and thus it would be possible to find rest for $S$, which would recover the inertial reference frames of SR.

The inverse transformations of space-time ($x^{\prime}\rightarrow X$) and ($t^{\prime}\rightarrow t$) in SSR for the special case $(1+1)D$ and also the general case $(3+1)D$ have already been explored in details in a previous paper\cite{N2016}. Thus, from such transformations above, we can obtain the following inverse transformation of velocity, namely: 

\begin{equation}
v_{rel}=v_{S'/S}=\frac{v_{S'/S_V}+v_{S/S_V}-V}{1+\frac{(v_{S'/S_V})(v_{S/S_V})}{c^2}-\frac{(v_{S'/S_V})V}{c^2}}. 
\end{equation}

The velocity transformation given by Eq.(14) leads to the following important cases:  

 {\bf a)} If $v^{\prime}=v_{S'}=v_S=v=V$, this implies in $``V+V"=V$. Once again we verify that the minimum speed $V$ is in fact invariant. 
 
 {\bf b)} If $v^{\prime}=v_{S'}=c$ (photon) and $v_S\leq c$, this leads to $v_{rel}=v_{S'/S}=c$. This just confirms that $c$ is invariant. 
 
 {\bf c)} If $v^{\prime}=v_{S'}>V$ and by considering $v_S=V$, this leads to $v_{rel}=v_{S'/S}=v_{S'}$. 
 
 From the case ({\bf c}), let us investigate the following specific cases, namely: 
 
 -$c_{1}$) If $v^{\prime}=v_{S'}=2V$ and by assuming that $v_S=V$, we would obtain $v_{rel}=v_{S'/S}=``2V+V"=2V$. 
 
 -$c_{2}$) If $v^{\prime}=v_{S'}=v_S=v$, this implies in $v_{rel}=v_{S'/S}=``v_S+v_S"=``v+v"=\frac{2v_S-V}{1+\frac{v_S^2}{c^2}(1-\frac{V}{v_S})}=\frac{2v-V}{1+\frac{v^2}{c^2}(1-\frac{V}{v})}$.
 
 In the Newtonian regime ($V<<v<<c$) for $c_2$ above, we recover the classical transformation, i.e., $v_{rel}=``v+v"=2v$. 
 
 In the relativistic regime ($v\rightarrow c$), we recover the Lorentz transformation of velocity given for this specific case $c_2$ ($v^{\prime}=v_{S'}=v_S=v$), i.e., we find 
 $v_{rel}=v_{S'/S}=``v_S+v_S"=``v+v"=2v_S/(1+v_S^2/c^2)=2v/(1+v^2/c^2)$.
 
 \section{\label{sec:level1} Equivalence of the SSR-metric with a dS-metric}

\begin{figure}
\begin{center}
\includegraphics[scale=0.27]{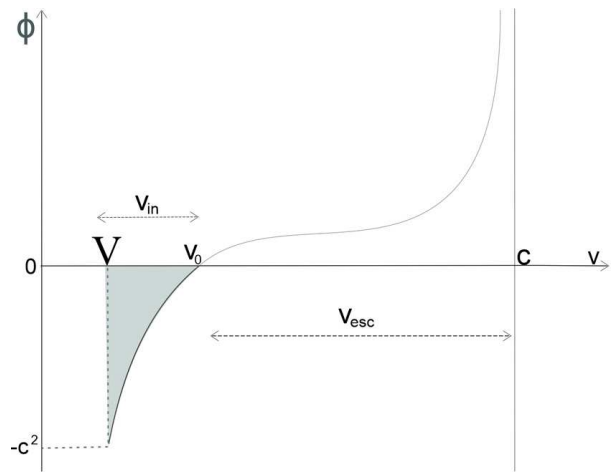}
\end{center}
\caption{This figure shows the scalar potential $\phi(v)=c^{2}\left(\sqrt{\frac{1-\frac{V^2}{v^2}}{1-\frac{v^2}{c^2}}}-1\right)$ given in function of speed [Eq.(16)]. It shows two phases, namely gravity (right side)/anti-gravity (left site), where the barrier at the right side represents the relativistic limit, i.e., speed of light $c$ with $\phi\rightarrow\infty$, and on the other hand, the barrier at the left side is the anti-gravitational limit only described by SSR with an invariant minimum speed $V$ associated with the potential $\phi_q(V)=\phi(V)=-c^2$. The intermediary region is the Newtonian regime ($V<<v<<c$), where occurs a phase transition of gravity/anti-gravity for $v=v_0=\sqrt{cV}$.}
\end{figure}
 
As the universal minimum speed related to the ultra-referential $S_V$ shoud be associated with the cosmological constant, let us show that the metric of SSR [Eq.(11)] is equivalent to a dS-metric, where there emerges a conformal factor depending on $\Lambda$\cite{Rodrigo}. To do that, let us use a toy model by considering a spherical universe with Hubble radius $R_H$ filled by a vacuum energy density $\rho$. 
 
According to this toy model, on the surface of the sphere by representing the frontier of the observable universe, all the objects like galaxies, etc experience an anti-gravity given by the accelerated expansion of the universe. This anti-gravitational effect is due to the whole vacuum energy (a dark mass) inside such a Hubble sphere. Thus, we think that each galaxy works like a proof body that interacts with this sphere having a dark mass $M_{\Lambda}(=M)$. Such interaction can be thought of as being the simple case of interaction between two bodies. In view of this, let us show that there is an anti-gravitational interaction between the ordinary proof mass $m_0$ (on the surface of the dark sphere) and the own dark sphere with a dark mass $M$. 

So, in order to investigate such anti-gravitational interaction between the proof mass $m_0$ and the dark mass $M$ of the Hubble sphere with radius $R_H$ in this toy model, let us first remind the model of a proof particle with mass $m_0$ that escapes from a gravitational potential $\phi$ on the surface of a certain sphere of matter with mass $M_{matter}$, namely $E=m_0c^2(1-v^2/c^2)^{-1/2}\equiv m_0c^2(1+\phi/c^2)$, where $E$ is the escape relativistic energy of the proof particle with mass
$m_0$ and $\phi=GM_{matter}/R$, $R$ being the radius of the sphere of matter. In this classical case, the interval of escape velocity ($0\leq v<c$) is associated with the interval of potential ($0\leq\phi<\infty$), where we define $\phi>0$ to be the well-known classical (attractive) gravitational potential. 

We should notice that the Lorentz symmetry violation in SSR is due to the presence of the ultra-referential $S_V$ (Fig.1) connected to the vacuum  energy that fills the dark sphere. Such energy has origin from a non-classical aspect of gravity that leads to a repulsive gravitational potential defined as being negative for representing anti-gravity, i.e., $\phi=\phi_q<0$ (Fig.2). 

In this toy model based on SSR-theory, we write the deformed relativistic energy of such a proof particle ($m_0$), as follows: 

\begin{equation}
E=m_0c^2\left(1+\frac{\phi}{c^2}\right)=
m_0c^2\left(\frac{{1-\frac{V^2}{v^2}}}{{1-\frac{v^2}{c^2}}}\right)^{\frac{1}{2}},
\end{equation}
from where we get
\begin{equation}
\phi=\phi(v)=\left[\left(\frac{{1-\frac{V^2}{v^2}}}{{1-\frac{v^2}{c^2}}}\right)^{\frac{1}{2}}-1\right]c^2. 
\end{equation} 

Here, we should realize that Eq.(16) reveals two situations, namely: 

i) the well-known Lorentz sector 
($\phi=(\gamma-1)c^2$) represents the gravity sector, since the sphere $M$ is composed by attractive (ordinary) matter. In this case, the speed $v$ is simply the escape velocity ($v_{esc}$), which is directed away from the sphere.

ii) the anti-gravity sector 
($\phi=\phi_q=(\theta-1)c^2$),
where $\theta=(1-V^2/v^2)^{1/2}$ is governed by a dark sphere with mass $M$. Here, the speed $v$ is the input speed ($v_{in}$) or the velocity of a proof particle that escapes from anti-gravity, i.e., $v(=v_{in})$ is directed into the sphere, since anti-gravity pushes the particle away. 

As SSR forbids rest of a particle according to Eq.(16), we must be careful to notice that $v$ cannot be zero even in the absence of potential $\phi$ ($\phi=0$), i.e, we find $v=v_0=\sqrt{cV}$, so that $\phi(v_0)=0$ in Eq.(16) (Fig.2). 

Due to the absence of gravitational potential 
($\phi=0$) at the point $v=v_0(\neq 0)$, this is the only velocity that means both of the escape and input velocities of a particle. Therefore, $v_0$ is a transition ``zero''-point between gravity and anti-gravity, which highlights the quantum nature of the space-time in SSR, thus leading to the uncertainty principle as shown in a previous paper\cite{uncertainty}. 

In short, from Eq.(16) and Fig.(2) we can see two regimes of gravitational potential, i.e., the classical (matter) and quantum (vacuum) regimes, namely: 

\begin{equation}
\phi=\phi(v)=\left\{
\begin{array}{ll}
\phi_{q}:&\mbox{$-c^2<\phi\leq 0$ for $V< v\leq v_0$}.\\\\
\phi_{m}:&\mbox{$0\leq\phi<\infty$ for $v_0\leq v<c$}, 
\end{array}
\right.
\end{equation}
so that the speed $v_0$ represents the point of transition ($\phi=0$) between gravity 
($\phi_{matter}=\phi_{m}=\phi>0$ for $v>v_0$) and anti-gravity when the vacuum governs 
($\phi_{quantum}=\phi_q=\phi<0$ for $v<v_0$). 

We must stress that $v_0$ is given with respect to the preferred reference frame $S_V$ (Fig.1). Therefore, it is an observer-independent velocity as well as any velocity $v$, which is given with respect to $S_V$, since $S_V$ is related to the unattainable minimum speed $V$, and thus there is no observer at the ultra-referential $S_V$.   

We realize that the most repulsive potential is 
$\phi=-c^2$, which is associated with the fundamental vacuum energy of the ultra-referential $S_V$ by imposing $v=v_{in}=V$ in Eq.(16), i.e., $\phi(V)=-c^2$ (Fig.2). 

So, by taking into account this model of a spherical universe with a Hubble radius $R_H(=R_u)$ and a vacuum energy density $\rho$, we obtain the total vacuum energy inside the sphere, i.e., $E_{dark}=\rho V_u=Mc^2$, where  $V_u$ is the spherical volume of the universe
and $M$ is the total dark mass associated with the vacuum energy inside the sphere. 

As the vacuum energy density $\rho$ is very low and the big sphere with Hubble radius 
$R_H(=R_u)$ presents a dark mass $M$, but having a very low dark mass density, then the Newtonian gravitational potential is a very good approximation that represents this toy model for the universe. So, in view of this, we get the following repulsive gravitational potential $\phi(=\phi_q<0)$ on the surface of such Hubble sphere (universe), namely: 

\begin{equation}
\phi=\phi_q=-\frac{GM}{R_u}=
-\frac{4\pi G\rho R_u^2}{3c^2}=-\frac{G\rho V_u}{R_uc^2},
\end{equation}
where $M=\rho V_u/c^2$, $\rho$ is the vacuum energy density and $V_u(=4\pi R_u^3/3)$ is the Hubble volume.

We already know that $\rho=\Lambda c^2/8\pi G$. So, by substituting this relationship in Eq.(18), we obtain the repulsive (quantum) potential, as follows: 

\begin{equation}
\phi=-\frac{\Lambda R_u^2}{6},
\end{equation}
where $R_u=R_H=cT_H(\sim 10^{26}$m) is the Hubble radius. $T_H\cong 13.7$ Gyear is the age of the universe (Hubble time). 

As the whole universe (a big sphere) is governed by the vacuum energy (a dark mass $M$), the speed $v$ in Eq.(16) is understood as the input speed $v_{in}$ in order to overcome the cosmological anti-gravity. Thus, the factor
$(1-V^2/v^2)^{1/2}$ [Eq.(16)] prevails for determining the potential $\phi$. In view of this, we will neglect the Lorentz factor $\gamma$ (attractive sector) in Eq.(16), and so we should consider only the repulsive sector ($V<v\leq v_0$) for obtaining the non-classical background potential $\phi(=\phi_q)$.

Furthermore, after neglecting $\gamma=
(1-v^2/c^2)^{-1/2}$ in Eq.(16), we will just compare its anti-gravity sector with Eq.(19) given for a certain radius $r(=ct)$, i.e., 
$\phi(=-\Lambda r^2/6)$, so that we find 
$\phi/c^2$, namely: 

\begin{equation}
\frac{\phi}{c^2}=-\frac{\Lambda r^2}{6c^2}=\left(1-\frac{V^2}{v^2}\right)^{\frac{1}{2}}-1, 
\end{equation}
where $\phi=\phi_q$, which represents the potentials of anti-gravity (Fig.2), 
being $0\leq\phi\leq-c^2$. 

By performing the calculations in Eq.(20), we rewrite the scale factor $\Theta(v)$ of the SSR-metric [Eq.(11)] in its equivalent forms, as follows:

\begin{equation}
\Theta(v)=\frac{1}{\left(1-\frac{V^2}{v^2}\right)}\equiv\frac{1}{\left(1+\frac{\phi_q}{c^2}\right)^2}\equiv\frac{1}{\left(1-\frac{\Lambda r^2}{6c^2}\right)^2}, 
\end{equation}
where we realize that there are three equivalent forms for representing $\Theta(v)\equiv\Theta(\phi_q)\equiv\Theta(\Lambda)$
as shown in Eq.(21). 

By replacing the factor $\Theta(v)$ of Eq.(11) (SSR-metric) by its equivalent form with dependence of $\Lambda$ shown in Eq.(21), we rewrite the background metric (SSR-metric) in its equivalent form within the dS-scenario, namely: 

\begin{equation}
d\mathcal S^{2}=\frac{1}{\left(1-\frac{{\Lambda}r^2}{6c^2}\right)^2}[c^2(dt)^2-(dx)^2-(dy)^2-(dz)^2], 
\end{equation}

or simply 

\begin{equation}
d\mathcal S^{2}=\Theta(\Lambda)\eta_{\mu\nu}dx^{\mu}dx^{\nu}, 
\end{equation}
where $\eta_{\mu\nu}$ is the Minkowski metric and
$\mathcal G_{\mu\nu}=\Theta(\Lambda)
\eta_{\mu\nu}$ is the SSR-metric with dependence of $\Lambda$. 

Of course if we make $\Lambda=0$ in Eq.(22), we get $\Theta=1$ and so we recover the Minkowski metric $\eta_{\mu\nu}$, where there is no cosmological constant and no anti-gravitational effect. In other words, as 
$\Lambda=-6\phi/r^2$ [Eq.(19)], 
for $r\rightarrow\infty$ 
($\Lambda\rightarrow 0$), the interval 
$d\mathcal S^{2}$ reduces to the 
Lorentz-invariant Minkowski interval $ds^2$, i.e., $d\mathcal S^{2}\rightarrow
ds^2=\eta_{\mu\nu}dx^{\mu}dx^{\nu}$
\cite{jgpereira}. 

We should realize that Eq.(22) represents a 
dS-metric which presents $\Lambda>0$, as we
must have $\phi<0$ (anti-gravity sector) according to Eq.(19).   

In view of Eq.(19) and Eq.(22), we first conclude that a cosmological constant $\Lambda$ emerges from SSR, i.e., $\Lambda=-6\phi/r^2$ [Eq.(19)]. We also conclude that there is a correspondance of SSR with the de-Sitter (dS)
relativity\cite{jgpereira} shown by 
Eq.(22) that is a dS-metric with a conformal 
factor\cite{jgpereira} given by 
$\Theta(\Lambda)$. 

We finally conclude that a small positive value of $\Lambda$ is plausible in the context of Eq.(19) and Eq.(22). So, in order to estimate the small order of magnitude of $\Lambda$, we first consider $\Lambda$ [Eq.(19)] given for the Hubble radius $R_H(\sim 10^{26})m$, so that we obtain 

\begin{equation}
\Lambda=\Lambda(R_H,\phi)=-\frac{6\phi}{R_H^2},
\end{equation}
where $r=R_H$ and $-c^2\leq\phi\leq 0$ (the shaded area in Fig.2). 

Finally, if we admit that the accelerated expansion of the universe is governed by the lowest potential $\phi=\phi(V)=-c^2$ associated with the fundamental vacuum energy at the ultra-referential $S_V$, we find 

\begin{equation}
\Lambda=\frac{6c^2}{R_H^2}\sim 10^{-35}s^{-2}. 
\end{equation}

Such a very small $\Lambda$ may have implication in a realistic cosmological scenario, however, more explorations are required to be done in that respect, which could be taken up as a future work. 

\section{\label{sec:level1} Conclusions and prospects} 

We have built an extension of Dirac's large number hypothesis (LNH) and so we have found a new constant of nature, namely a universal minimum speed ($V\sim 10^{-14}$m/s). Such speed
is a new kinematic invariant given for lower energies, which led to a new Deformed Special 
Relativity (DSR) so-called Symmetrical Special
Relativity (SSR), from where there emerged
the cosmological constant, which allowed us 
to show the equivalence of SSR-metric with a kind of dS-metric. The small value of the 
cosmological constant was estimated. 

We have shown that the Bohr velocity ($v_{B}=e^2/\hbar\sim 10^{5}$m/s) in the fundamental state of the Hydrogen atom, the Bohr radius ($a_0$) of the orbit of an electron in its ground state and the fine structure constant 
($\alpha$) depend on the Eddginton number $N$. 

The investigation of symmetries of the SSR-theory should be done by means of their association with a new kind of electromagnetism when we are in the limit $v\rightarrow V$, which could explain the problem of high magnetic fields in magnetars\cite{gravastar}\cite{gravastar1}\cite{Greiner}, super-fluids in the interior of gravastars\cite{gravastar1} and other types of black hole mimickers\cite{haw}\cite{Tds,Tds1,haw2}.\\

\end{document}